\documentclass{aa}
\usepackage{graphicx}
\usepackage[varg]{txfonts}
\usepackage{cases}
\usepackage{natbib}

\begin{document}
   \title{Decayless longitudinal oscillations of a solar filament maintained by quasi-periodic jets}

   \author{Y. W. Ni\inst{1,2}, J. H. Guo\inst{1,2}, Q. M. Zhang\inst{3}, J. L. Chen\inst{3}, C. Fang\inst{1,2} and P. F. Chen\inst{1,2}}

   \institute{School of Astronomy and Space Science, Nanjing University, Nanjing 210023, China \\
              \email{chenpf@nju.edu.cn; y.w.ni@smail.nju.edu.cn}
              \and
              Key Laboratory of Modern Astronomy and Astrophysics (Nanjing University), Ministry of Education, Nanjing 210023, China
              \and
              Key Laboratory of Dark Matter and Space Astronomy, Purple Mountain Observatory, CAS, Nanjing 210033, China \\
              \email{zhangqm@pmo.ac.cn}
              }

   \date{Received; accepted}
    \titlerunning{Decayless filament oscillations maintained by quasi-periodic jets}
    \authorrunning{Ni et al.}

 \abstract
{As a ubiquitous phenomenon, large-amplitude longitudinal filament oscillations usually decay in 1--4 periods. Recently, we observed a decayless case of such oscillations in the corona.}
{We try to understand the physical process that maintains the decayless oscillation of the filament.}
{Multi-wavelength imaging observations and magnetograms are collected to study the dynamics of the filament oscillation and its associated phenomena. To explain the decayless oscillations, we also perform one-dimensional hydrodynamic numerical simulations using the MPI-AMRVAC code.}
{In observations, the filament oscillates decaylessly with a period of $36.4 \pm 0.3$ min for almost 4 hours before eruption. During oscillations, four quasi-periodic jets emanate from a magnetic cancellation site near the filament. The time interval between neighboring jets is $\sim 68.9 \pm 1.0$ min. Numerical simulations constrained by the observations reproduced the decayless longitudinal oscillations. However, it is surprising to find that the period of the decayless oscillations is not consistent with the pendulum model.}
{We propose that the decayless longitudinal oscillations of the filament are maintained by quasi-periodic jets, which is verified by the hydrodynamic simulations. More importantly, it is found that, when driven by quasi-periodic jets, the period of the filament longitudinal oscillations depends also on the driving period of the jets, not simply the pendulum period. With a parameter survey in simulations, we derived a formula, by which one can derive the pendulum oscillation period using the observed period of decayless filament oscillations and the driving periods of jets.}


 \maketitle

\section{Introduction}\label{sect1}
Solar prominences are bright cloud-like structures composed of plasmas with lower temperature and higher density compared to the ambient corona \citep[e.g.,][]{Lab10, mackay10, parenti14, vial15, gib18}. The typical plasma density of prominences is approximately $10^{10}$--$10^{11}$ cm$^{-3}$, and the typical temperature is approximately $10^4$ K. When a prominence rotates from the solar limb to the solar disk, it is called a filament, typically visible at H$\alpha$, \ion{He}{I} 10830 {\AA}, EUV and radio wavelengths \citep[e.g.,][]{vanBall04, ber10, schmi10, schmi14, shen15, yan15, yangl17}. Except some filaments that are maintained by chromospheric siphon flows \citep[e.g.,][]{wang99} so that magnetic dips are not necessary \citep{karp01} for them, many filaments are believed to be supported by magnetic dips, which allow filament threads to be held in stable equilibrium with the upward Lorentz force balancing the gravity \citep{chen20}. The corresponding magnetic configurations with dips include sheared arcades \citep{kipp57}, magnetic flux ropes \citep{kupe74}, or their combination  \citep{guo10, liu12}.

A filament might exist for up to several days or months, during which it is continually disturbed by sporadic travelling waves, microflares, cold surges and hot jets, and ever-lasting convection on the solar surface. Once disturbed, a filament would begin to oscillate. Filament oscillations can be classified into two categories (i.e., transverse oscillations \citep{her11, dai12, shen14b, zqm18} and longitudinal oscillations \citep{jing03,  luna14, shen14a, maz20}). For both categories, observations indicated that filament oscillations generally damp out in 1--4 periods \citep{arre18} because of the existence of various dissipation mechanisms such as 
radiative losses, thermal conduction, resonant absorption, and wave leakage. However, \citet{chen08} investigated the dynamics of a prominence with spectroscopic observations, and found that the prominence oscillations lasted for 4 hours without damping. Note that the prominence oscillated with two periods: 20 min and $\sim$60 min. Presumably, the shorter period corresponds to the transverse mode, and the longer period corresponds to the longitudinal mode. Since such a decayless oscillation was followed by eruption, they tentatively proposed that long-duration oscillations could be a precursor of prominence eruptions and coronal mass ejections (CMEs). Since then, an increasing number of observations have indicated that decayless oscillations of prominences/filaments are followed by eruptions \citep{gosa09, duch10, LiT12, mash16, mash19}. More recently, \citet{luna18} conducted a survey of 196 filament longitudinal oscillation events in the first half of 2014 and found that while the distribution of the ratio between the decay time and the period peaks at 1.25 and most events damp out in 3 periods, a few events manifest long-term oscillations without significant damping or even with amplification. They also noticed that such events are followed by eruptions.

These observations raise two important questions: What is the mechanism to maintain the undamped or even amplified longitudinal oscillations in some filaments? What is the relationship between the decayless prominence oscillations and ensuing prominence eruptions? In this paper, we study an umdamped event of filament longitudinal oscillation on 2014 July 5 in order to answer these two questions. The observations and data analysis are shown in Sect.~\ref{s-data}. The observational results are described in Sect.~\ref{s-res}. In Sect.~\ref{s-num}, one-dimensional (1D) hydrodynamic simulations are performed to reproduce the observations, which is followed by discussions in Sect.~\ref{s-disc}. The main conclusions are summarized in Sect.~\ref{s-sum}.

\section{Observations and data analysis} \label{s-data}

\begin{figure*}
\includegraphics[width=18cm,clip]{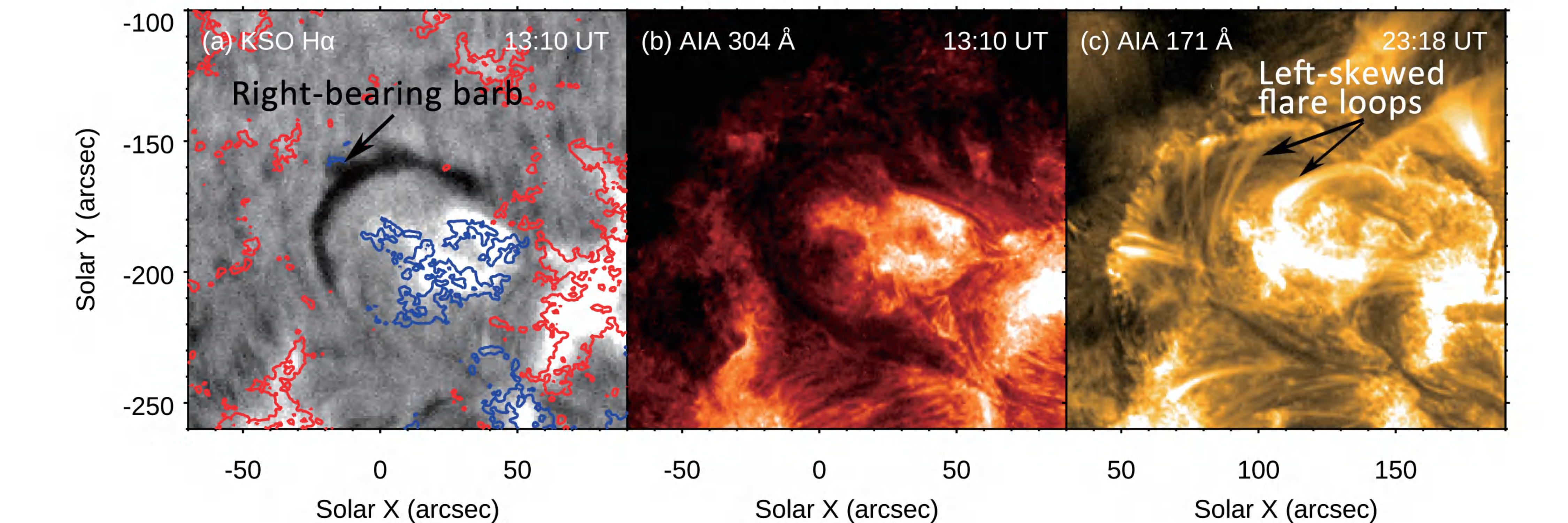}
\centering
\caption{Panel (a): The corresponding H$\alpha$ image of the filament provided by the Kanzelh{\"o}he Observatory at 13:10:46 UT, where the red and blue contour lines represent different levels of magnetic field strength ranging from $-50$ G to $50$ G. 
The arrow indicates a right-bearing barb of the filament.
Panel (b): The SDO/AIA 304 \AA\ image observed at 13:10:43 UT.
Panel (c): The SDO/AIA 171 \AA\ image observed at 23:19:36 UT. 
\textbf{The arrows mark left-skewed flare loops formed after the filament eruption.}}
\label{fig1}
\end{figure*}

The left panel of Fig.~\ref{fig1} shows the local H$\alpha$ image around the filament, which is located near active region (AR) 12104 on 2014 July 5. The H$\alpha$ data are observed by the Kanzelh{\"o}he Observatory. The line-of-sight magnetogram provided by the Helioseismic and Magnetic Imager (HMI) on board SDO is overplotted on the H$\alpha$ image, where the red and blue solid curves mark the contours of negative and positive magnetic field strength at different levels ranging from $-50$ G to $50$ G. The filament spine is clearly distinguished and roughly distributed along the polarity inversion line \citep{martin90}. Since $\sim$18:30 UT on July 5, the filament starts to oscillate. After oscillating for almost 4 hours, the filament erupts, leaving behind a C3.8-class solar flare, which is registered by the GOES satellite  \citep{cjl21}. No CME is clearly identified in the Large Angle and Spectrometric COronagraph (LASCO) coronagraph observations, but a faint CME is detected by the COR2 coronagraph onboard the Solar Terrestrial Relations Observatory (STEREO). The discrepancy probably results from the unfavored observing vantage of LASCO since the source active region is close to the solar disk center in the LASCO field of view, whereas it is close to the limb in the STEREO field of view.

The KSO H$\alpha$ telescope has a spatial resolution of $\sim$2{\arcsec} with a 2k $\times$ 2k CCD camera in operation \citep{steine00}, which is relatively low in revealing the fine structures of the oscillating filament. 
The evolution of the filament is also observed at multiple EUV wavelengths by the Atmospheric Imaging Assembly \citep[AIA; ][]{Lem12} onboard the Solar Dynamics Observatory (SDO). 
The AIA data have a spatial resolution of $\sim$1$\farcs$2 and a cadence of 12 s. 
In this paper, only the 171\AA\ and 304 \AA\ images are used. The Helioseismic and Magnetic Imager \citep[HMI; ][]{scher12} onboard SDO provides the photospheric magnetograms, which can trace the associated evolution of the magnetic field on the solar surface. 
The AIA and HMI data are calibrated by utilizing the standard Solar Software (SSW) routines \textit{aia\_prep.pro} and \textit{hmi\_prep.pro}.
The full-disk H$\alpha$ and AIA 304 {\AA} images are coaligned using the cross-correlation method, and all images are de-rotated to a pre-oscillation time, that is, 18:35:12 UT. The right panel of Fig. \ref{fig1} depicts the SDO/AIA 304 {\AA} image taken at approximately the same time as the left panel. The EUV filament has a longer extension in the southern part, which is barely visible in H$\alpha$, a typical feature of filaments \citep{aula02}. 

\section{Observational results} \label{s-res}

\begin{figure*}
\includegraphics[width=12cm,clip]{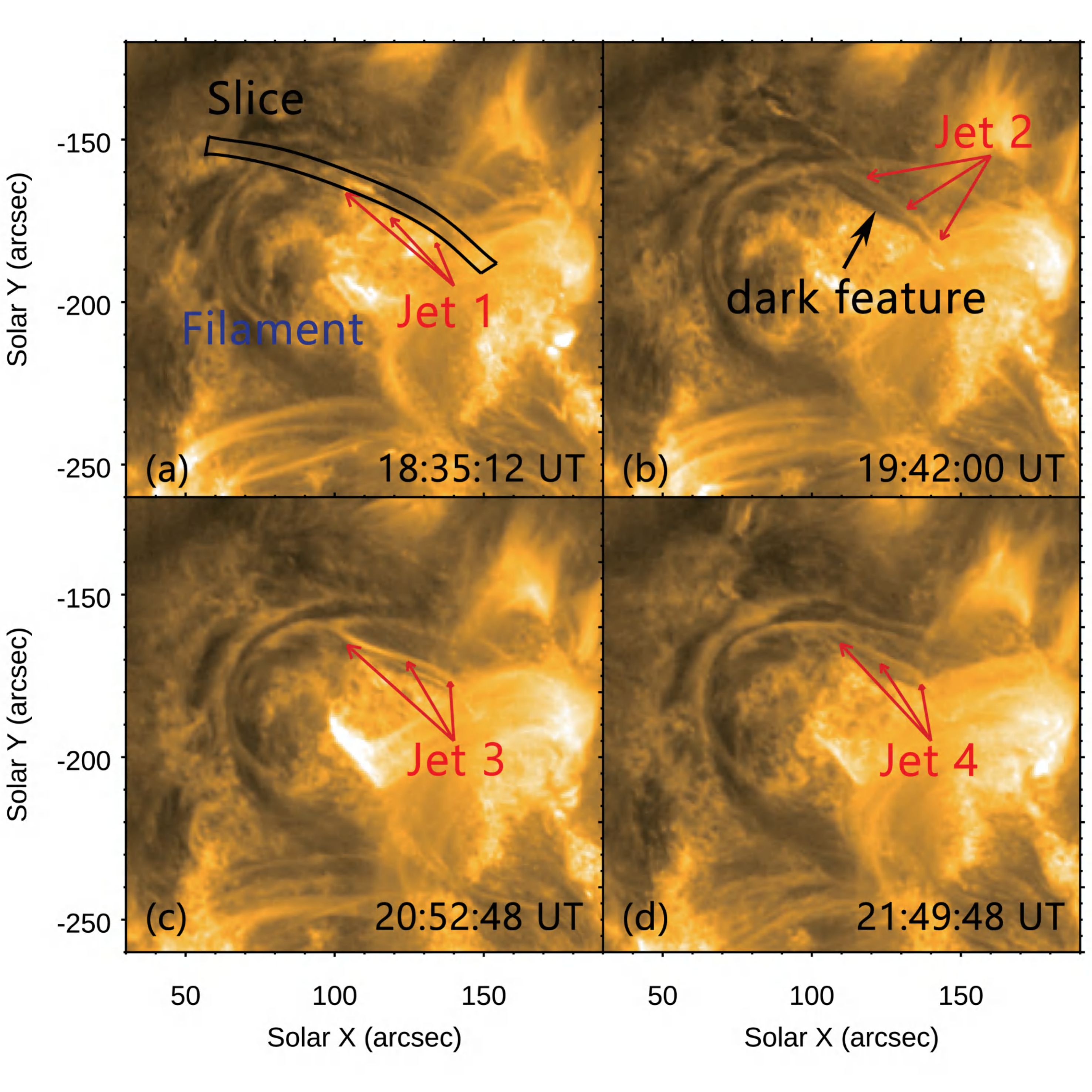}
\centering
\caption{Temporal evolution of the AR filament in AIA 171 {\AA} from ~18:35 UT to ~22:20 UT. In panel (a), a slice is chosen in order to plot the time-distance diagram in Fig. \ref{fig3}.
The black arrow in panel (b) marks dark features accompanying the jets.
The animation of this figure is available online.}
\label{fig2}
\end{figure*}

The filament has a quite long lifetime, existing in AR12104 from June 29 as a prominence above the east solar limb to July 5, when it is near the solar disk center. At $\sim$18:35 on July 5, the first jet emanates from nearby, moving toward the filament at a speed of 100.1 km s$^{-1}$. The jet is manifested as bright emissions in 171 \AA, and is accompanied by dark features, as displayed by panel (b) of Fig.~\ref{fig2}, which depicts several snapshots of the filament evolution observed by SDO/AIA at 171 \AA. In response to the impact of the jet, the filament starts to oscillate along the direction of its threads, therefore, such an oscillation is a typical longitudinal oscillation. The second and third jets emanate from the same site at 19:42 UT and 20:52 UT, respectively, making the filament oscillate with an almost constant amplitude. When the brightest fourth jet is ejected at 21:49 UT, the filament starts to erupt, leaving flaring loops in the low corona and flaring ribbons near the solar surface.

\begin{figure*}
\includegraphics[width=14cm,clip]{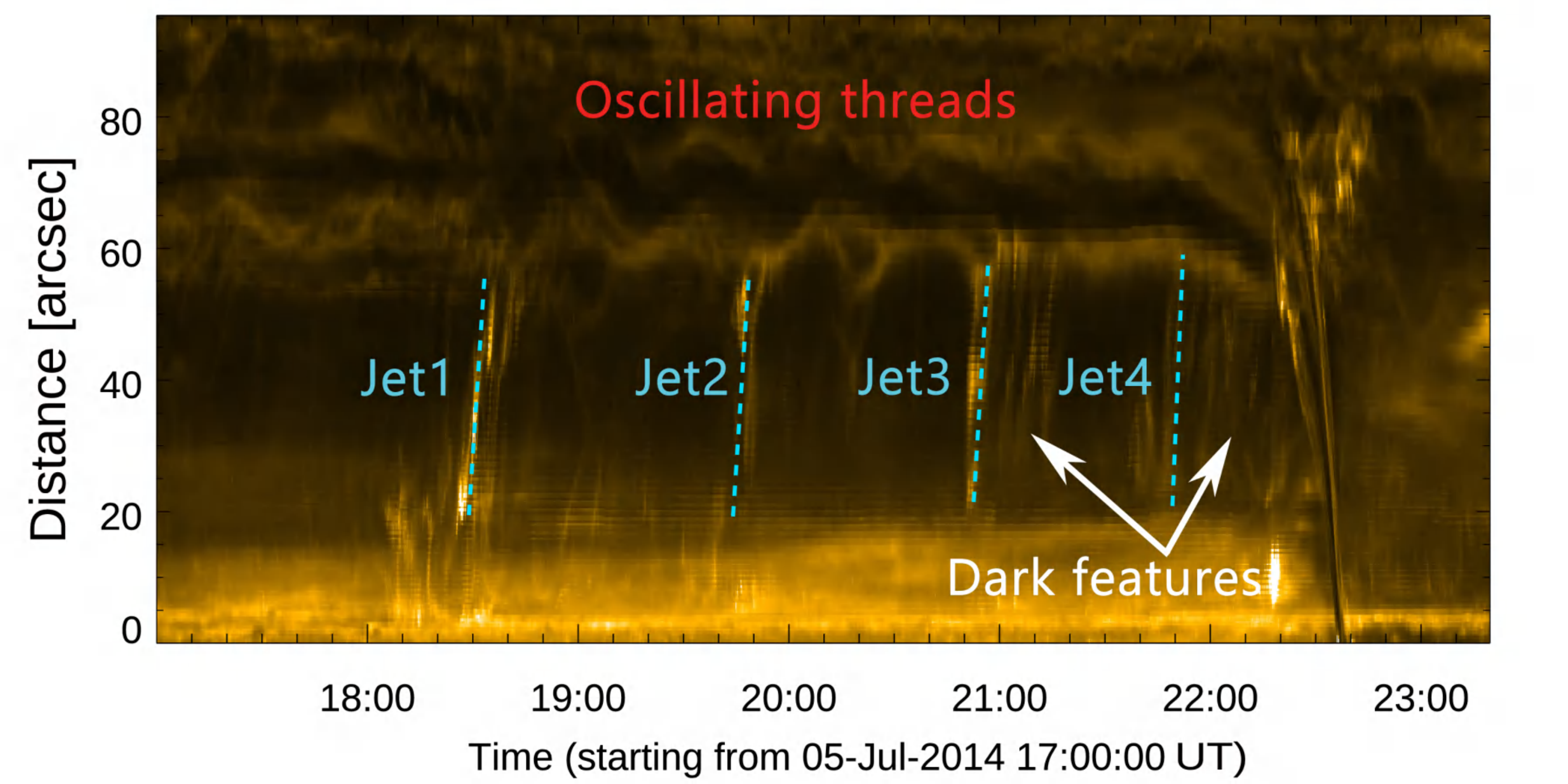}
\centering
\caption{Time-distance diagram of the AIA 171 \AA\ intensity along the slice marked in Fig. \ref{fig2}(a). The distance increases from west to east, therefore the dark ridge with a strong slope around 22:40 UT means the filament was erupting toward west. The figure indicates decayless filament oscillations associated with quasi-periodic jets.}
\label{fig3}
\end{figure*}

To see the sequence of the jets and the resulting filament oscillations, we select a slice along the trajectory of the jets and along the filament threads, as bounded by the solid line in Fig. \ref{fig2}(a). Note that the positive direction of the slice is chosen from west to east so that the jet propagation is along the positive direction. In this case, the filament eruption, which is toward the west, would appear along the negative direction of the slice. The time-distance diagram of the 171 \AA\ intensity distribution along the slice is plotted in Fig. \ref{fig3}, where the filament is located at a distance of 70\arcsec\ in Fig. \ref{fig3}. Quasi-periodic jets are identified as the bright ridges extending from the position at $\sim$20\arcsec\ in Fig. \ref{fig3} to the edge of the filament, which are indicated by the cyan dashed lines. The slopes of the four bright ridges characterize the moving velocities of the jets, which are found to be 100.1, 100.1, 108.0 and 157.5 km s$^{-1}$, respectively. The time interval between the first and the second jets is approximately 67.9 min, that between the second and third jets is approximately 69.9 min, and that between the third and fourth jets is approximately 58.5 min.
It is shown that the first and second time intervals are close to each other, and the third interval is slightly smaller, indicating that the jets are quasi-periodic. The period averaged by the first two intervals is 68.9 min. A fifth jet starts at 22:20 UT. However, it is not related to the filament oscillations since the filament has already risen rapidly. The properties of the quasi-periodic jets are listed in Table \ref{tab1}. Besides, the duration of each jet is $\sim$5 min. It should be noted that Fig. \ref{fig3} does not only show 4 quasi-periodic jets before filament eruption, but also reveals continuous dark ridges along the trajectory with similar velocities. These absorptive ridges are supposed to be cold surges, and they appear more frequently, even between successive hot jets.

\begin{table}
\caption{Parameters of the quasi-periodic jets observed by SDO/AIA.}
\label{tab1}
\centering
\begin{tabular}{cccc}
\hline\hline
Event  & $t_{\rm start}$ (UT) & $t_{\rm arrival}$ (UT) & Speed (km s$^{-1}$)\\
\hline
Jet 1...... & 18:35:13 & 18:39:59 & 100.1 \\
Jet 2...... & 19:42:12 & 19:46:54 & 100.1 \\
Jet 3...... & 20:52:00 & 20:56:52 & 108.0 \\
Jet 4...... & 21:49:48 & 21:54:23 & 157.5 \\
\hline
\end{tabular}
\end{table}

\begin{figure}  
\includegraphics[width=8cm,clip]{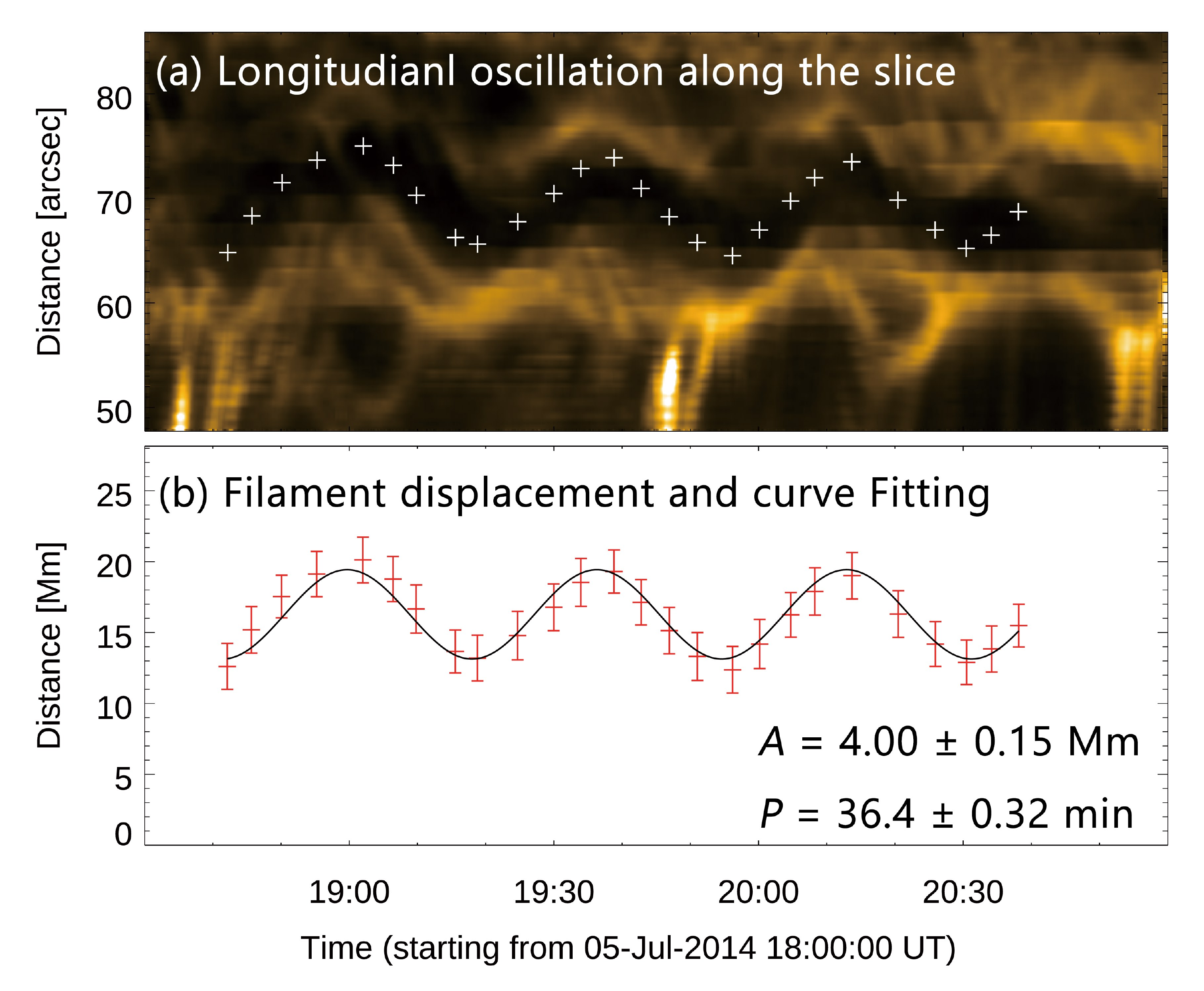}
\centering
\caption{Panel (a): Close-up view of Fig. \ref{fig3} around the filament, highlighting the filament oscillations. Panel (b): Evolution of the filament centroid (red error bars) and the fitted curve (black line).}
\label{fig4}
\end{figure}

To quantitatively investigate the oscillatory behavior of the filament, we zoom-in Fig. \ref{fig3} around the filament part, and the corresponding time-distance diagram is displayed in Fig. \ref{fig4}(a). We find that the oscillatory behavior of the filament threads differs significantly from the ordinary large-amplitude longitudinal oscillations which damp out in $\sim$3 periods \citep{luna18,daij21}. The filament exhibits a decayless oscillation. To calculate the position of the filament thread at each time, we trace the upper and lower boundaries of the filament thread by the strongest intensity gradient along the slice direction since the dark filament is often bounded by a bright layer. We define the middle point of the upper and lower boundaries to be the centroid of the filament. The filament intensity distribution along the slice is fitted with a Gaussian profile in order to obtain the full width at half maximum (FWMH) absorption of radiation, which is then taken as the uncertainty of the centroid position \citep{awa19}. The evolution of the filament centroid is displayed in Figure \ref{fig4}(b). We use a decaying sine function of time to fit the temporal variation of the filament thread centroid, that is, 

\begin{equation} \label{eqn-1}
  y=y_0+A_0\sin(\frac{2\pi}{P}t+\phi_0)e^{-t/\tau},
\end{equation}
\noindent
where $y_0$ and $\phi_0$ denote the initial position and phase angle of the oscillation, respectively. $A_0$, $P$ and $\tau$ denote the initial amplitude, period, and damping time of the large amplitude filament oscillation, respectively. The evolution of the filament longitudinal displacement with uncertainty is fitted via the Markov Chain Monte Carlo method \citep{sharma17} with one hundred fitting tests. As a result, the initial amplitude of filament oscillation is $A_0=4.0\pm0.15$ Mm, the period is $36.4\pm0.3$ min, and the damping time is 10$^{13}$ min, which implies that the oscillation is identical to a decayless one. According to the pendulum model \citep{luna12, zqm12}, the period $P$ is mainly related to the curvature radius $R$ of the dipped magnetic field line by $P=2\pi \sqrt{R/g_{\odot}}$, where $g_{\odot}=2.74\times 10^2$ m s$^{-2}$ is the gravitational acceleration near the solar surface. The curvature radius of the magnetic dip is estimated to be $33.3\pm0.5$ Mm.

\begin{figure*}  
\includegraphics[width=12cm,clip]{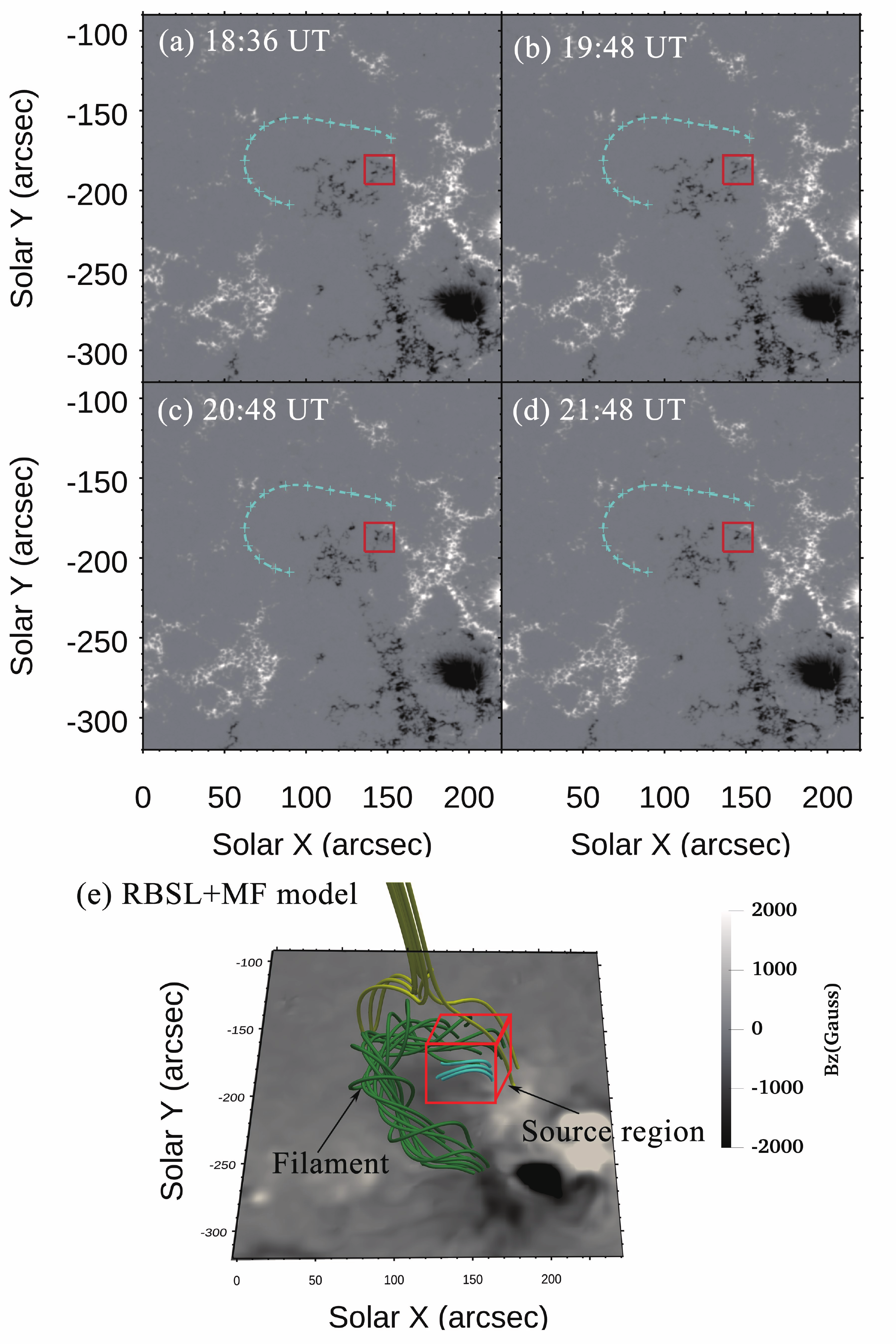}
\centering
\caption{Panels (a)--(d): Four snapshots of the SDO/HMI magnetogram around the filament channel, where white/black means positive/negative polarity, and the red square boxes correspond to the jet source site. The filament spine is represented by the cyan dashed lines.
Panel(e): Coronal magnetic field reconstructed using the RBSL flux rope and magneto-frictional model. The green field lines represent the flux rope and the yellow lines indicate that the overlying magnetic field contains a fan-spine structure.
The red cube corresponds to the red square boxes in the top four panels, where magnetic reconnection probably happened.}
\label{fig5}
\end{figure*}

From Fig. \ref{fig2} and the associated animation, it is seen that the repetitive jets originate from the western side of the filament channel. After comparing the AIA 171 \AA\ images and the HMI longitudinal magnetograms, we identify the location of the jet source in the HMI magnetogram, which is marked by the small red square boxes in Fig. \ref{fig5}. Figure \ref{fig5}(a--d) display the photospheric magnetograms at 18:36, 19:48, 20:48 and 21:48 UT. We find that inside the red boxes the positive and negative magnetic polarities approach each other, and magnetic cancellation occurs.

To understand the magnetic configuration of the jet formation, the coronal nonlinear force-free field (NLFFF) is reconstructed by using the magneto-frictional model \citep{gy16a, gy16b}, where a flux rope is embedded in a potential field in advance \citep{vanBall00, vanBall04}.
The regularized Biot–Savart laws (RBSL) method ensures the embedded flux rope to be in internal equilibrium beforehand \citep{titov18, torok18, gy19, gjh21}. 
Then the magnetic field is relaxed to a force-free state. After 60,000 steps in our magneto-friction code, the force-free metric, $\Sigma ({\mathbf J}\times {\mathbf B})_i/(JB)$, drops to approximately 0.3, indicating that the magnetic field approaches the force-free state, where ${\mathbf B}$ is the magnetic field, and ${\mathbf J}$ is the current density.

Figure \ref{fig5}(e) displays the coronal magnetic field overplotted on the photospheric magnetogram around the filament channel. It is seen that the magnetic field lines in cyan are strongly sheared compared to the twisted field lines of the filament in green. Magnetic reconnection is expected to happen between the two parts. After such interchange reconnection, jets would propagate along the twisted magnetic field lines, and hit the filament threads.

\section{Numerical simulations} \label{s-num}

Without extra energy supply, an oscillating filament would damp out in $\sim$3 periods even under the effects of radiation and thermal conduction. According to the previous section, we argue that the decayless oscillations of the filament on 2014 July 5 are due to the impact of quasi-periodic jets, which push the filament threads repetitively. To confirm this conjecture, we perform 1D hydrodynamic numerical simulations as in our previous works \citet{xia11, zqm12, zqm13, zqm20a, zyh17}.

The whole procedure consists of four steps: (1) Filament thread formation: A filament thread is formed in a prescribed 1D magnetic field via the chromospheric evaporation and coronal condensation mechanism; (2) relaxation: After halting the localized heating at the footpoints, the filament thread naturally relaxes to a quasi-static equilibrium state; (3) perturbations: four episodes of impulsive heating are imposed at one footpoint of the magnetic field line, which excite high-speed jets hitting the filament, resulting in filament longitudinal oscillations. The details are presented as follows.

\subsection{Simulation setup}

As described by \citet{xia11}, we numerically solve the following 1D hydrodynamic equations of single-fluid fully ionized plasma:

\begin{equation} \label{eqn-2}
  \frac{\partial \rho}{\partial t}+\frac{\partial}{\partial s}(\rho v)=0 \,,
\end{equation}

\begin{equation} \label{eqn-3}
  \frac{\partial}{\partial t}(\rho v)+\frac{\partial}{\partial s}(\rho v^2+p)=\rho g_{\parallel}(s) \,,
\end{equation}

\begin{equation} \label{eqn-4}
  \frac{\partial \varepsilon}{\partial t}+\frac{\partial}{\partial s}(\varepsilon v+pv)=\rho g_{\parallel}v+H-n_{\rm H}n_{\rm e}\Lambda(T)+\frac{\partial}{\partial s}(\kappa \frac{\partial T}{\partial s}) \,,
\end{equation}
\noindent
where $\rho$, $T$ and $v$ are the basic physical parameters (i.e., plasma density, temperature, and velocity). The plasma density is related to the number density of protons by $\rho=1.4 m_{\rm p} n_{\rm H}$, when helium is considered in the solar corona. Correspondingly, the plasma pressure is formulated as $p=2.3 n_{\rm H} k_{\rm B} T$. For the mono-atomic ideal gas, the specific heat ratio is $\gamma=5/3$, hence the internal energy is described as $\varepsilon= \rho v^2 /2 + p/(\gamma -1)$. In the above equations, $g_\parallel(s)$ is the component of gravity at a distance $s$ along the flux tube. In addition, $H(s)$ is set as the local heating, and $\Lambda(T)$ is the optically-thin radiation loss function. Spitzer heat conductivity is taken to be $\kappa=10^{-6} T^{5/2}$ ergs cm$^{-1}$ s$^{-1}$ K$^{-1}$. The above equations are numerically solved with the open-source code, MPI-AMRVAC\footnote{http://amrvac.org} \citep{kep12, xia18}. The numerical scheme \emph{HLL} is adopted as the Riemann solver and we use a 3-level mesh refinement, reaching the highest spatial resolution of $40.5$ km.

As mentioned in Section \ref{sect1} and \citet{chen20}, magnetic dips exist in many filaments, although not all, as implied by the prevalence of filament longitudinal oscillations. For the filament in this study, on the one hand, longitudinal oscillations are observed, which implies that the filament is supported by magnetic dips. On the other hand, the filament has right-bearing barbs, and the flaring loops are left-skewed relative to the magnetic neutral line, implying negative helicity of the magnetic system. According to the method proposed by \citet{chen14}, the filament is identified to be supported by a magnetic flux rope, rather than a sheared arcade. Hence, magnetic dips indeed exist in this filament. As illustrated in Fig.~\ref{fig6}, we study the motion of the plasma along a dipped magnetic flux tube that is composed of the following parts:
(1) a vertical segment extending to a height of $s_1=18$ Mm from the each footpoint to the corona;
(2) two symmetrical quarter circles with a length of $s_2=7.48$ Mm as the shoulders of the dipped magnetic field line;
(3) in the middle part, there is a straight uniform helix, which is expressed as follows:

\begin{figure*}
\includegraphics[width=14cm,clip]{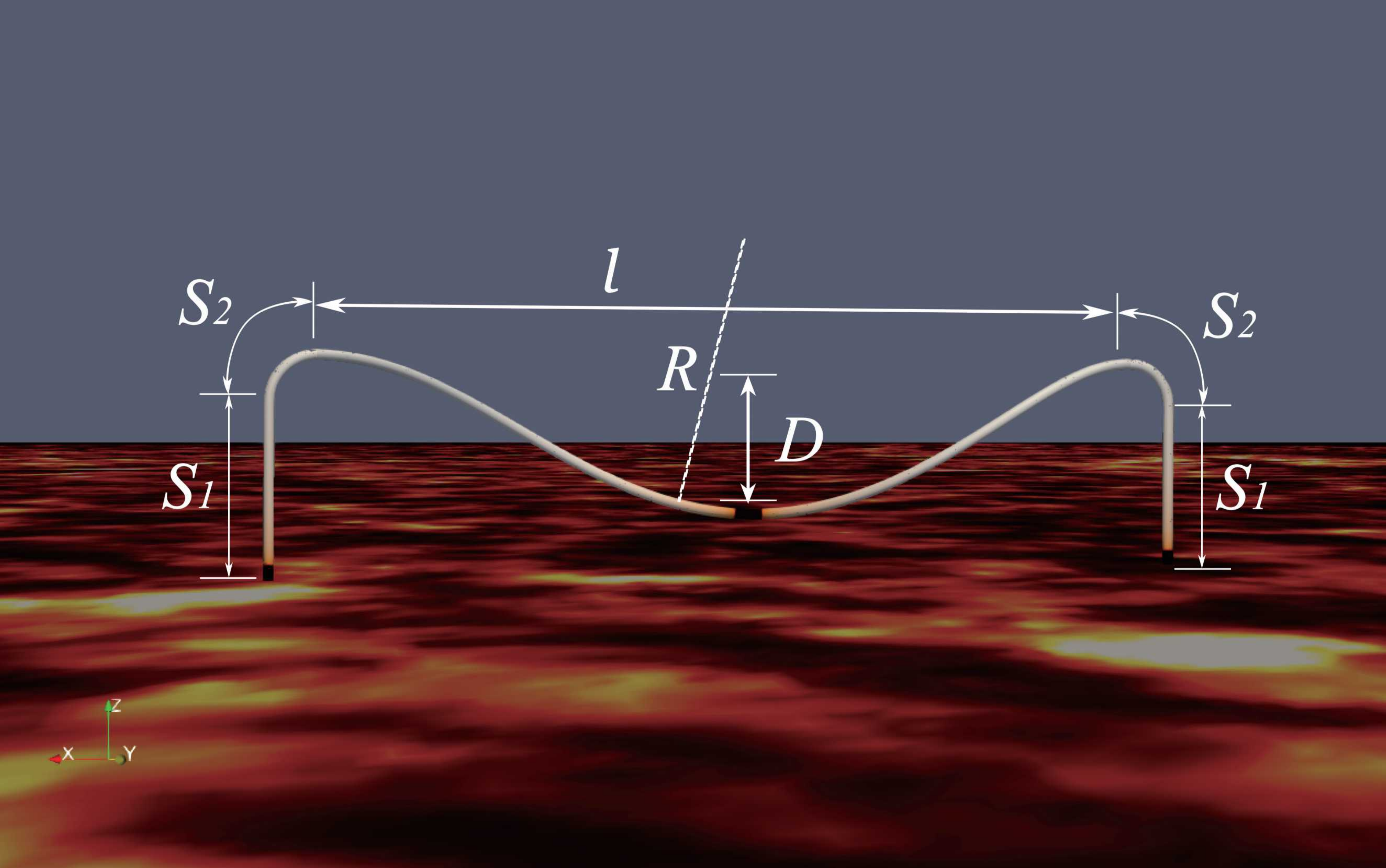}
\centering
\caption{The geometry of the flux tube in our simulations, where the central part is a helix described by Equation (\ref{eqn-6}).
The natural coordinate is along the tube starting from the left-side footpoint.}
\label{fig6}
\end{figure*}

\begin{equation}\label{eqn-6}
\left\{
             \begin{array}{lr}
             x=\theta l/2\pi , &  \\
             y=0.5D\sin\theta, & 0\leq \theta\leq 2\pi, \\
             z=0.5D\cos\theta, &
             \end{array}
\right.
\end{equation}
 where $l$ is the axis length of the helix and $D$ is the radius of the magnetic flux tube, which is also equal to the dip depth. In this paper, we set $l=90.5$ Mm and $D=16.6$ Mm, so that the resulting curvature radius of the magnetic dip would be $R=33.3$ Mm, the same as derived from the observations in the previous section based on the pendulum model.

A steady background heating $H_b=E_0e^{-(s-s_{min})/\lambda}+E_0e^{-(s_{max}-s)/\lambda}$ is imposed to maintain the hot corona. Localized heating is then imposed at the two footpoints symmetrically so as to form a filament thread near the magnetic dip. After the localized heating is halted, the filament thread is relaxed for 5.5 hr in order to reach a quasi-static equilibrium state. Since we adopted the optically-thin radiation loss for the whole dynamics of the filament from formation to oscillation, the core temperature of the filament remains at $\sim$17,000 K, which is consistent with previous works \citep[e.g.,][]{karp06}. In this case, the plasma is fully ionized in our simulated filaments. In reality, the core temperature of filaments is several thousand Kelvin, and the plasma is partially ionized, with the ionization rate being about 0.3 \citep{Lab10}. Fortunately, even in the partially-ionized situation, it was verified that a single-fluid assumption as used in this paper is still valid \citep{terr15}.

To mimic the periodic jets, four heating pulses are then applied successively at the left footpoint of the magnetic flux tube with the intervals being the same as in the observations, which are listed in Table \ref{tab1}. Similar to \citet{zqm20a}, the impulsive heating $H_i$ is expressed as follows:
\begin{equation} \label{eqn-5}
  H_i(s)=E_1 \exp{\left[{-\frac{(s-s_{\rm peak})^2}{s_{\rm width}^2}-\frac{(t-t_{\rm peak})^2}{t_{\rm scale}^2}}\right]},
\end{equation}
\noindent
where the spatial distribution is the same for various jets with $s_{peak}=6$ Mm and $s_{width}=3$ Mm, determined from the AIA 171\AA\ images. The damping time of the impulsive heating is 5 min for all the four thermal pulses. All pulses are set with the impulsive heating rate $E_1=4.4\times10^{-3}$ erg cm$^{-3}$ s$^{-1}$.

To investigate the effects taken by the periodic thermal pulses, we also set up a control case, where only the first thermal pulse is introduced to cause the filament to oscillate. Its results will be compared to those in the case with multiple pulses.

\subsection{Simulation results}

\begin{figure*}
\includegraphics[width=16cm,clip]{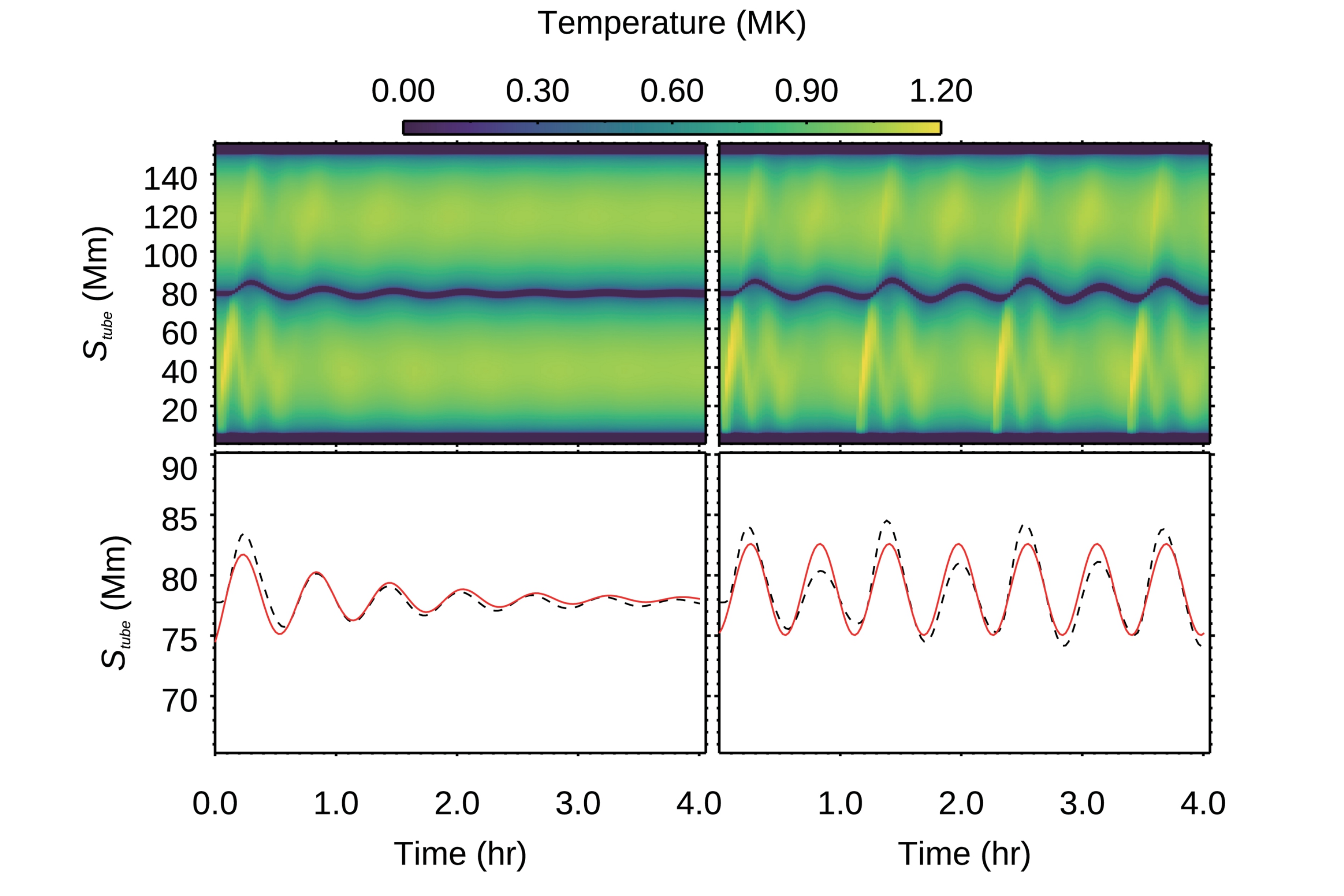}
\centering
\caption{Top row: Time-distance diagrams of the temperature distribution along the magnetic flux tube in the control case (left) and in the multi-pulse case (right). Bottom row: Temporal evolution of the displacement of the filament thread center in the control case (left) and in the multi-pulse case (right), where the black dashed lines correspond to the simulation results and the red solid lines correspond to the fitting curves.}
\label{fig7}
\end{figure*}

The top row of Fig.~\ref{fig7} displays the evolution of the temperature distributions in the control case (top-left panel) and in the multi-pulse case (top-right panel), where the blue segment in the middle around $s$=80 Mm corresponds to the filament thread. 
In the control case the filament thread begins to oscillate after being impinged by a single jet. The oscillation damps away quickly as expected. On the contrary, in the multi-pulse case the filament thread oscillates continually. Once the oscillation is to damp out, a second jet comes in, so the oscillation is amplified again.

In order to quantitatively compare the two cases, we extract the position of the filament thread center at each time in each case, and its evolution is displayed as black dashed lines in the bottom row of Fig. \ref{fig7}, where the bottom-left panel corresponds to the control case, whereas the bottom-right panel corresponds to the multi-pulse case. 
The displacement evolution is fitted with a decaying sine function $y=y_0+A_0 e^{-t/\tau}\sin(2\pi t/P+\phi_0)$, so that we can derive all the oscillation parameters, including the oscillation period ($P$) and decay time ($\tau$).
With the least-square method, we fit the simulation results with the red solid lines in the bottom row of Fig. \ref{fig7}. The period is 36.4 min in the control case and 34.35 min in the multi-pulse case, and the damping time is 72.7 min in the control case and $5.2\times10^{8}$ min (almost infinity) in the multi-pulse case. The fitting results are listed in Table \ref{tab2}. Surprisingly it is found that the periods are different between the control case and the multi-pulse case, and only in the control case, the oscillation period is consistent with the pendulum model.

\begin{table}
\caption{Parameters of simulated filament oscillations in the control case and the multi-pulse case.}
\label{tab2}
\centering
\begin{tabular}{ccccc}
\hline\hline
Case & $A_0$ (Mm) & $P$ (min) & $\tau$ (min) & $\tau/P$ \\
\hline
Control case...... & 6.51 & 36.4 & 72.7 & 2.0 \\
\hline
Multi-pulse case...... & 3.91 & 34.35 & $\infty$ & $\infty$ \\
\hline
\end{tabular}
\end{table}

\section{Discussions} \label{s-disc}

\subsection{How filament decayless oscillations are maintained}

Filament oscillations are a ubiquitous phenomenon in the solar atmosphere. With the existence of radiative cooling, heat conduction, and other dissipation processes, such oscillations generally damp out within $\sim$3 periods. However, filament oscillations are occasionally decayless for a long time, which was proposed to be one of the precursors for filament eruptions or CMEs \citep{chen08}. The decayless behavior implies that extra energy must be supplied to the filament somehow. In this paper, we revealed that filament decayless oscillations are accompanied by quasi-periodic jets. Our 1D hydrodynamic simulations confirmed that quasi-periodic jets from the low corona hit the filament thread, maintaining the decayless longitudinal oscillations of the filament.

\begin{figure}
\includegraphics[width=8cm,clip]{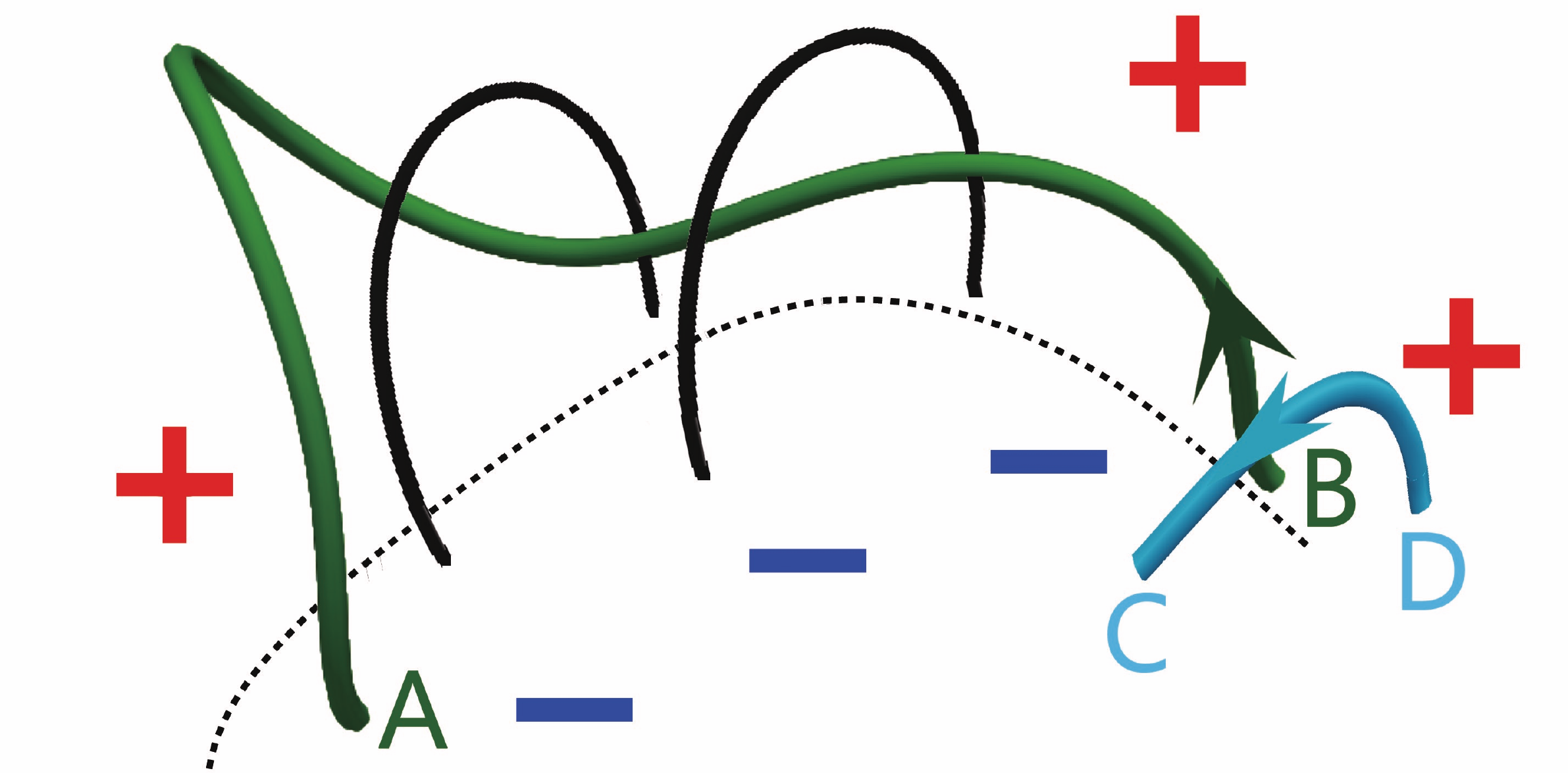}
\centering
\caption{Schematic sketch of the magnetic reconnection process, which triggers both repetitive jets and the onset of the filament eruption. The green line AB represents the magnetic field supporting the filament, and the cyan line CD represents the neighboring strongly-sheared field line. The interchange reconnection between lines AB and CD leads to jets and the expansion of the post-reconnection field line AD.}
\label{fig8}
\end{figure}

Solar jets are prevalent, which can be observed in solar active regions \citep{shibata92, shibata94, ylh11, chd17, shen17}, quiet-Sun regions and even coronal holes \citep{chd12, tian14}, with typical velocities in the range of 30--300 km s$^{-1}$. They have been well explained to be due to magnetic reconnection either in the low corona or in the chromosphere. As illustrated in Fig. \ref{fig5}, the EUV jets in this paper indeed originate from a magnetic cancellation site near the eastern end of the filament channel. The approaching and canceling positive and negative magnetic polarities imply that the reconnection occurs in the low solar atmosphere, e.g., the upper chromosphere. This is consistent with the co-existence of bright and dark EUV jets, which means that if the reconnection occurred in the low corona, only hot and bright EUV jets would be visible. Based on the photospheric magnetogram and the EUV images showing coronal loops and jet trajectories, the magnetic structure and connectivity are sketched in Fig. \ref{fig8}, where the dashed line corresponds to the magnetic neutral line, the black lines are the envelope magnetic field lines overlying the filament, the green line represents the magnetic field lines supporting the filament, and the cyan line corresponds to the low-lying strongly sheared field lines. The spatial relationship between the green line and the cyan line is derived from the extrapolated coronal magnetic configuration with the same color coding in Fig. \ref{fig5}(e).

As footpoints B and C are dragged to approach each other by the photospheric convection flows, interchange reconnection is triggered between field lines AB and CD. After reconnection, the green field line is rooted at point D with a concave-upward kink structure above footpoint D. While the smaller reconnected loop BC sinks down, producing magnetic cancellation phenomenon, the upward reconnection outflow forms an EUV jet, moving along the post-reconnection field line AD. A bundle of magnetic field lines are involved in the reconnection process. For the field lines threading the filament, the jet directly hits the filament to oscillate longitudinally, as indicated by observations \citep{luna14, zqm17b, zqm20c} and illustrated by numerical simulations in this paper in 1D or in \citet{luna21} in 2D. For the field lines encircling the filament, the jet plasma does not impact the filament material directly. Instead, the magnetic rearrangement after reconnection would impose a pulse on the filament, and the filament would begin to oscillate as well. Such a process can be simulated only in 2 dimensions \citet{luna21} or 3 dimensions.

The above-mentioned jet-induced filament oscillations can be well fit into the emerging flux trigger mechanism of CMEs \citep{chen00, kusa12}: As illustrated in Fig. \ref{fig8}, the filament system is initially hindered from erupting under the magnetic tension force of the black field lines. After the interchange reconnection between field lines AB and CD, magnetic loop AB becomes loop AD, with a concave-upward kink structure around the reconnection site. Such a kink structure provides an upward Lorentz force, which pulls up the field line AD, leading to the rise motion of the filament and the overlying field lines. As described in \citet{chen00}, a current sheet below the rising flux rope is formed, whose reconnection leads to the main flare and the filament eruption. The only difference between Fig. \ref{fig8} and the emerging flux trigger mechanism \citep{chen00, kusa12} is that the magnetic loop CD has a pre-existing field, rather than emerging from the subsurface. Similar to \citet{chen08}, we argue that, as the triggering progresses, the interchange reconnection illustrated in Fig. \ref{fig8} is intermittent, rather than continuous, as indicated by the quasi-periodic jets. What determines the $\sim$68.9 min quasi-period of the reconnection (hence the jets) remains an interesting question noteworthy to be explored \citep{ChenJ15, Cheung15, Para20}.

\subsection{Relationship between the filament oscillation period and the jet period}

Although we successively reproduced the filament decayless longitudinal oscillations in the numerical simulations as quasi-periodic jets hit the filament, we obtained an unexpected result: While the oscillation period of the damped oscillation triggered by a single pulse is consistent with that predicted by the pendulum model \citep{luna12, zqm13}, the fitted period of the multi-pulse driven oscillations is approximately 10\% smaller, although the curvature radius of the magnetic dip is the same as in the single-pulse case. This implies that the fitted period of the driven oscillations of the filament is not the pendulum oscillation itself, and it also depends on the driving period of the impinging jets $P_{jet}$.

To investigate how the fitted period of the driven oscillations in observations ($P_{obs}$) varies with the driving period of the jets ($P_{jet}$), we conduct a parameter survey, where 19 other cases with different $P_{jet}$ are simulated with the 1D hydrodynamic equations (\ref{eqn-2}--\ref{eqn-4}). In all these cases, the curvature radius of the magnetic dip is kept the same, and that is $R=33.9$ Mm, which means that the intrinsic period of the pendulum model, $P_{pendu}$, is 36.4 min.

\begin{figure*}
\includegraphics[width=16cm,clip]{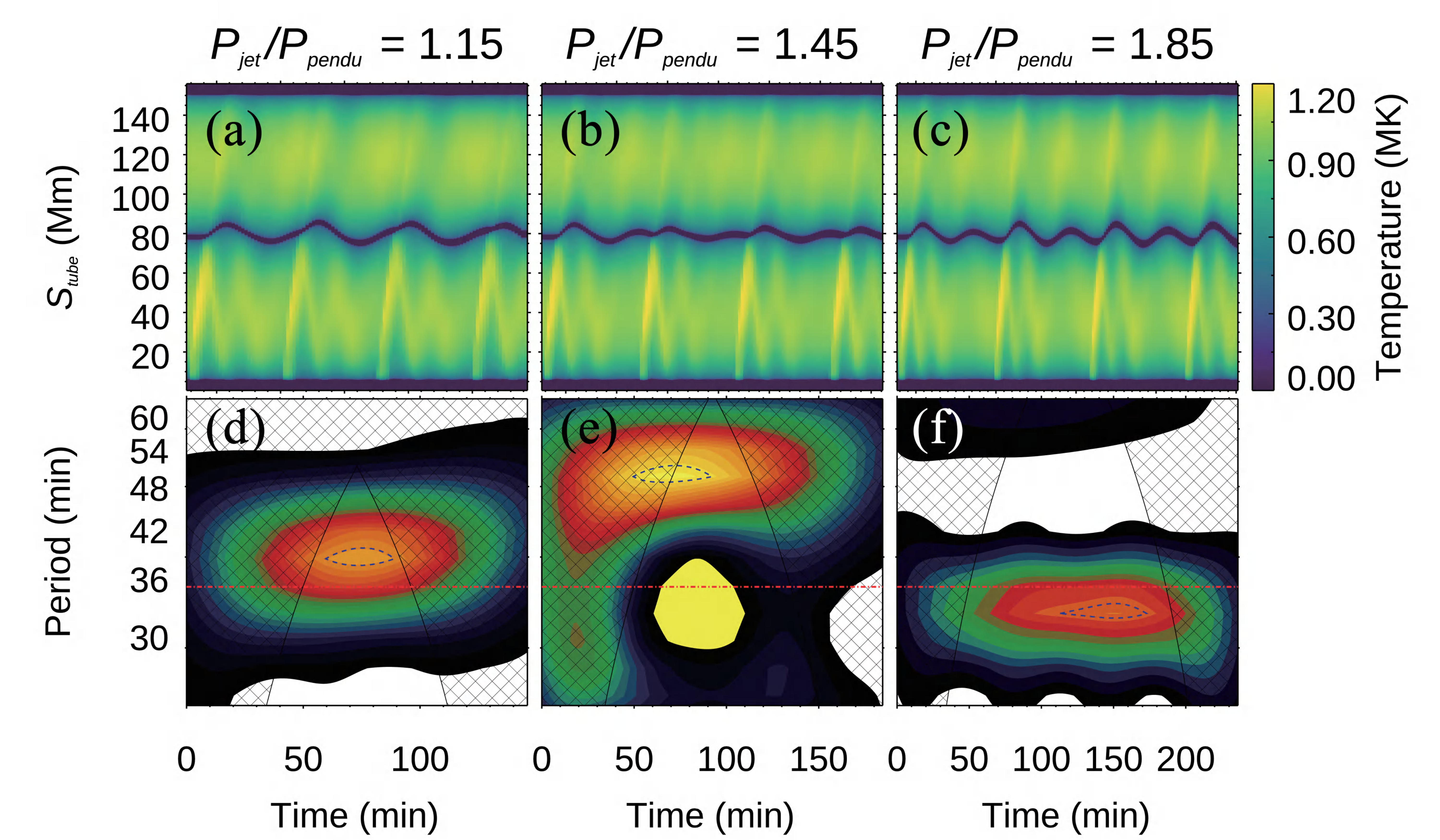}
\centering
\caption{Top row: Time-distance diagrams of the temperature distributions in three cases when the driving period of the jets is $1.15 P_{pendu}$, $1.45 P_{pendu}$, and $1.85 P_{pendu}$, respectively. Bottom row: The corresponding wavelet periodograms of the three cases, where the black dashed lines correspond to the significance of 95\%, and the horizontal red line marks the pendulum period}, $P_{pendu}$.
\label{fig9}
\end{figure*}

It is found that when $P_{jet}$ is close to $n P_{pendu}$ (where $n$=1, 2, ...), the driven oscillation of the filament is more like a decayless oscillation. However, when $P_{jet}$ is close to $(n+0.5) P_{pendu}$ (where $n$=1, 2, ...), the filament oscillation becomes chaotic. The top row of Figure \ref{fig9} displays the time-distance diagrams of the temperature distribution in three typical cases where $P_{jet}$ is $1.15 P_{pendu}$, $1.45 P_{pendu}$, and $1.85 P_{pendu}$, respectively. The bottom panels show the corresponding wavelet periodogram of the filament displacement in each case. When $P_{jet}=1.15 P_{pendu}$, Fig. \ref{fig9}(a) indicates that soon after the filament overpasses the central position, it is further pushed by a second jet. As a result, the oscillation period indicated by panel (d) becomes $P_{obs}=39.74$ min, which is larger than the pendulum period $P_{pendu}=36.48$ min. This can also be understood as follows: After the filament overpasses the central position, the restoring force (i.e., the gravity) is backward. The impact of a second jet acts as an outward force, which reduces the restoring force. A weaker restoring force results in a longer oscillation period. On the contrary,  when $P_{jet}=1.85 P_{pendu}$, Fig. \ref{fig9}(c) indicates that before the filament approaches the central position, a second jet pushes it from the back, enhancing the restoring force. As a result, the oscillation period becomes shorter (i.e., $P_{obs}=0.94 P_{pendu}$). When $P_{jet}=1.45 P_{pendu}$, Fig. \ref{fig9}(b) indicates that the jet propulsion is almost out of phase with the oscillating filament. As a result, the oscillation deviates from being quasi-periodic and becomes chaotic. In its wavelet periodogram in Fig. \ref{fig9}(e), we can still see its main period, which is actually the driving period of the jets.

\begin{figure}
\includegraphics[width=8cm,clip]{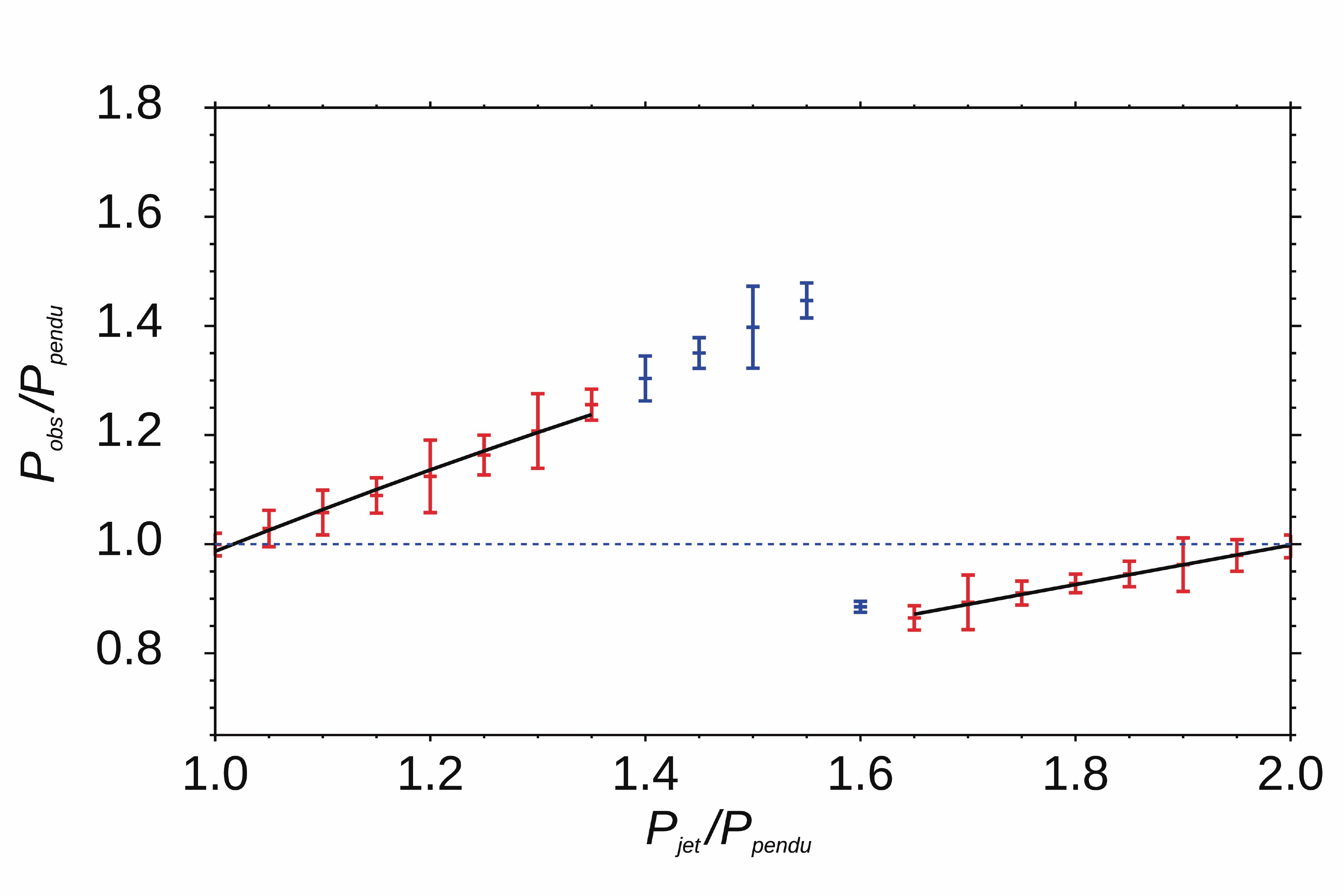}
\centering
\caption{Dependence of the observed period ($P_{obs}$) on the jet period $P_{jet}$ in our parameter survey, where both quantities are normalized by $P_{pendu}$. The blue dashed line marks the intrinsic period determined by the pendulum model.}
\label{fig10}
\end{figure}

For all the 20 cases, we perform a wavelet analysis of the evolution of the filament displacement to derive the ``observed'' oscillation period ($P_{obs}$). Its variation with the jet period $P_{obs}$ is shown in Fig. \ref{fig10}, where both parameters are normalized by the pendulum period $P_{pendu}$. It is found that $P_{obs}$ indeed deviates from the intrinsic period of the pendulum model ($P_{pendu}$) when the filament is driven by periodic jets, and the deviation depends on the driving period of the jet, $P_{jet}$. It is seen that, as $P_{jet}$ increases from $P_{pendu}$ to 2$P_{pendu}$, the ``observed'' oscillation period ($P_{obs}$) first increases to more than 1.4$P_{pendu}$, then drastically drops down to 0.9$P_{pendu}$, and then gradually increases back to $P_{pendu}$. Excluding the non-periodic cases when $P_{jet}$ is $\sim$1.5$P_{pendu}$, whose data points are color-coded in blue, all other cases (red data points) display quasi-period oscillations. We fit the red data points with the following formula:

\begin{equation}
\frac{P_{obs}}{P_{pendu}}=\left\{
\begin{array}{ll}
a(\frac{P_{jet}}{P_{pendu}})^2+b\frac{P_{jet}}{P_{pendu}}, &  1.00 \leq P_{jet}/{P_{pendu} \leq 1.35};\\
c\frac{P_{jet}}{P_{pendu}}+ d, & 1.65 \leq P_{jet}/P_{pendu} \leq 2.0.
\end{array}\right.
\label{eq-c}
\end{equation}

With the least-square method, the fitted parameters are $a=-0.20$, $b=1.20$, $c=0.36$, and $d=0.28$. In real observations, the pendulum period $P_{pendu}$ is an unknown. Therefore, we rewrite Eq. (\ref{eq-c}) in the following way:

\begin{equation}
P_{pendu}=\left\{
\begin{array}{ll}
0.20P_{jet}^2/(1.20P_{jet}-P_{obs}), & {1.00 \leq P_{jet}/P_{obs} \leq 1.07};\\
3.62P_{obs}-0.36P_{jet},  & {1.88 \leq P_{jet}/P_{obs} \leq 2.00.}
\end{array} \right.
\label{eq-d}
\end{equation}

With this formula, we can easily determine the pendulum period $P_{pendu}$ (hence the curvature radius) of the magnetic dips once we have measured the decayless oscillation period ($P_{obs}$) and the driving period of jets $P_{jet}$.

Back to the 2014 July 5 event, the observed period of the filament decayless oscillation is $P_{obs}\sim 36.4$ min, and the driving period of the jets is $P_{jet}\sim 68.9 \pm 1.0$ min. According to our Equation (\ref{eq-d}), the pendulum oscillation period $P_{pendu}$ should be 40.2 min, which corresponds to a curvature radius of 40.2 Mm. As a test, we perform another hydrodynamic simulation, with the parameters in Equation (\ref{eqn-5}) being $l=109.5$ Mm and $D=20.2$ Mm in order to match the curvature radius of $R=40.2$ Mm. With the quasi-periodic jets impinging the filament, the filament starts to oscillate decaylessly. It is found that the period of the decayless oscillations is $36.14 \pm 0.62$ min, which is almost the same as in observations.

It is also noticed that, although the fitted period of the decayless oscillations is not the intrinsic period of the pendulum model, the initial stage of the oscillation before the second jet hits the filament should be identical to the single-pulse case and hence should possess the pendulum period. Therefore, rather than fitting the whole evolution in Fig. \ref{fig4}, we fit the oscillation during the interval of 18:30--19:40 UT, and it is revealed that the fitted period is $40.0 \pm 0.24$ min, which is very close to the derived pendulum period from Equation (\ref{eq-d}).

\section{Summary} \label{s-sum}
In this paper, we used SDO/AIA and KSO H$\alpha$ data to analyze the decayless longitudinal oscillations of the filament in active region AR12104 on 2014 July 5. The main results are summarized as follows:
\begin{enumerate}
\item{The observations reveal that quasi-periodic homologous jets drive the large-amplitude decayless longitudinal oscillation of the filament before eruption.
    Our 1D hydrodynamic numerical simulations verified the causal relationship.}

\item{The decayless filament oscillations were followed by the filament eruption and a CME. All the observational features can be fit into the big picture of the CME triggering process described in \citet{chen00} and \citet{chen08}: Interchange magnetic reconnection between the magnetic field of the filament system and the ambient results in two consequences: On the one hand, quasi-periodic jets are produced, which hit the filament, leading to decayless longitudinal oscillations. On the other hand, the post-reconnection large-scale field has a concave-upward kink, whose Lorentz force pulls the filament system to rise, triggering the onset of filament eruption. The ensuing magnetic reconnection below the filament makes the final eruption as described by the standard CME/flare model. Our data analysis and simulations validated the physical ground of the decayless filament oscillations as a precursor of CMEs.}

\item{When driven by quasi-periodic jets or other perturbations, the period of the filament longitudinal oscillations, $P_{obs}$, deviates from the intrinsic period of the pendulum model ($P_{pendu}$) up to 40\% depending on the driving period of the jets, $P_{jet}$. One cannot use the pendulum model to derive the curvature radius directly. With a parameter survey, we propose an empirical formula to relate $P_{obs}$ and $P_{jet}$ to $P_{pendu}$ (i.e., Eq. (\ref{eq-d})). Alternatively, if we fit the oscillations in the initial stage before the second jet hits the filament, the fitted period is also the pendulum period}, which can be used to derive the curvature radius of the magnetic dip.
\end{enumerate}

It is noted that the filament plasma in our simulations was considered to be fully-ionized for simplicity. In real situations the temperature of the filament plasma is $\sim$7000 K, where partial ionization and optically thick radiation begin to take effect for internal waves propagating inside the filament \citep{Ball18, Ball20, Ball21}. For the global pendulum mode oscillations studied in this paper, these internal waves are not directly involved. Still, considering partial ionization and optically thick radiation would increase the density of the filament thread, which might prolong the decay time for any episode of filament longitudinal oscillations. Such an effect will be studied in future papers.

\begin{acknowledgements}
The authors thank the anonymous referee for constructive suggestions, and SDO and GONG teams for the open data policy. The research is funded by the National Key Research and Development Program of China (2020YFC2201200), NSFC (11961131002, 11773079, and 11533005), and the Strategic Priority Research Program on Space Science, CAS (XDA15052200, XDA15320301).
\end{acknowledgements}

\bibliographystyle{aa}
\bibliography{reference}

\end{document}